\documentclass{moriond}

\def\be{\begin{equation}}
\def\ee{\end{equation}}
\def\bea{\begin{eqnarray}}
\def\eea{\end{eqnarray}}

\newcommand{\Photo}{}

\begin{document}
\vspace*{4cm}
\title{TESTING GENERAL RELATIVITY WITH BLACK HOLE X-RAY DATA: RECENT PROGRESS AND FUTURE DEVELOPMENTS}

\author{ COSIMO BAMBI }

\address{Center for Field Theory and Particle Physics and Department of Physics\\Fudan University, 200438 Shanghai, China}

\maketitle\abstracts{
The theory of General Relativity has successfully passed a large number of observational tests. The theory has been extensively tested in the weak-field regime with experiments in the Solar System and observations of binary pulsars. The past five years have seen significant advancements in the study of the strong-field regime, which can now be tested with gravitational waves, X-ray data, and mm Very Long Baseline Interferometry observations. Here I summarize the state-of-the-art of the tests of General Relativity with black hole X-ray data, discussing its recent progress and future developments.}


\section{Introduction}

The theory of General Relativity was proposed by Einstein at the end of 1915. After more than 100~years and without any modification, it is still our standard framework for the description of gravitational fields and the spacetime structure. The theory has been extensively tested, but mainly in the weak field regime~\cite{Will:2014kxa}. In the past five years, there have been significant advancements~\cite{Bambi:2015kza} and we can now test the strong gravity region around black holes with gravitational waves~\cite{Yunes:2016jcc}, X-ray data~\cite{Cao:2017kdq}, and mm Very Long Baseline Interferometry observations~\cite{Psaltis:2020lvx}.

\begin{figure}[t]
\centerline{\includegraphics[width=0.6\linewidth]{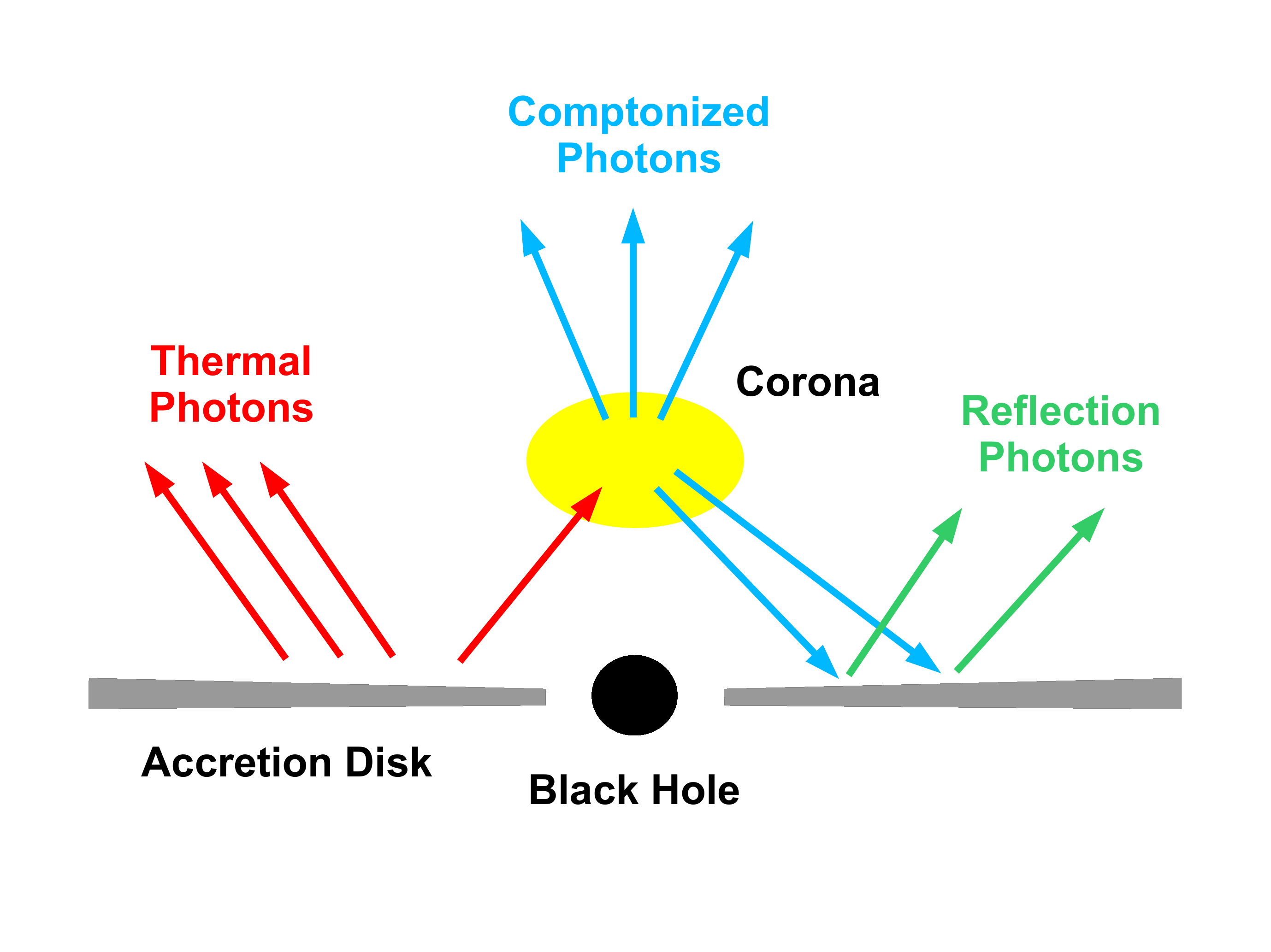}}
\vspace{-1.0cm}
\caption[]{Cartoon of the disk-corona model.}
\label{f-corona}
\end{figure}

Fig.~\ref{f-corona} shows the astrophysical system for testing General Relativity with black hole X-ray data~\cite{Bambi:2020jpe}. A black hole is surrounded by a geometrically thin and optically thick accretion disk. The gas in the disk is in local thermal equilibrium: every point on the surface of the disk emits a blackbody-like spectrum and the whole disk has a multi-temperature blackbody spectrum. The emission from the inner part of the accretion disk is peaked in the soft X-ray band (0.1-1~keV) for stellar-mass black holes ($M \approx 10~M_\odot$) and in the UV band (1-100~eV) for supermassive black holes ($M \sim 10^5$-$10^{10}~M_\odot$). The ``corona'' is some hotter ($\sim 100$~keV) gas around the black hole and the inner part of the accretion disk. It may be the atmosphere above the accretion disk, the accretion flow in the plunging region between the inner edge of the disk and the black hole, or the base of the jet, and more than one corona may coexist at the same time. Thermal photons of the accretion disk inverse Compton scatter off free electrons in the corona, generating a continuum with a power-law spectrum and a high-energy cutoff. A fraction of the Comptonized photons illuminate the disk: Compton scattering and absorption followed by fluorescent emission produce a reflection spectrum.

The reflection spectrum of the disk is characterized by fluorescent emission lines below 10~keV, in particular the iron K$\alpha$ complex at 6.4~keV for neutral or weakly ionized iron and up to 6.97~keV for H-like iron ions, and a Compton hump peaked at 20-30~keV. The iron K$\alpha$ complex is a very prominent and intrinsically narrow feature, so it is particularly suitable for measuring relativistic effects occurring in the strong gravity region around the black hole (gravitational redshift, Doppler boosting, and light bending)~\cite{Fabian:1989ej,Laor:1991nc,Bambi:2017khi}. The analysis of the thermal spectrum of the disk can also provide some valuable information about the spacetime geometry, because the disk's temperature increases as we approach to the black hole and from the shape of the spectrum it is possible to infer some properties of the innermost stable circular orbit of the spacetime~\cite{Zhang:1997dy,Kong:2014wha}.


\section{Testing the Kerr black hole hypothesis}

The spacetime metric around astrophysical black holes is thought to be described well by the Kerr solution of General Relativity~\cite{Bambi:2017khi}. However, macroscopic deviations from the Kerr geometry can be expected in other theories of gravity, in scenarios with large quantum gravity effects, or in the presence of some exotic matter fields. All these scenarios of new physics motivate current efforts of testing the spacetime around astrophysical black holes and check whether its geometry is consistent with the Kerr solution.

In the past few years, my group at Fudan University has developed the relativistic reflection model {\tt relxill\_nk}~\cite{rnk1,rnk2,rnk3,rnk4} and the multi-temperature blackbody model {\tt nkbb}~\cite{nkbb1,nkbb2}. Both models are specifically designed to test the Kerr black hole hypothesis from the analysis of relativistic reflection features and disk's thermal spectrum, respectively.

Following an agnostic approach, the spacetime metric around a black hole can be described by a deformed Kerr metric in which {\it ad hoc} deformation parameters are introduced to quantify possible deviations from the Kerr solution. {\tt relxill\_nk} and {\tt nkbb} calculate reflection and thermal spectra in such a deformed Kerr spacetime. From the analysis of X-ray data of specific sources, we can constrain the value of these deformation parameters and verify the Kerr black hole hypothesis.

One of the most popular metrics for testing the Kerr black hole hypothesis is the Johannsen metric~\cite{Johannsen:2015pca}. Its line element reads
\begin{eqnarray}\label{eq-jm}
ds^2 &=&-\frac{\tilde{\Sigma}\left(\Delta-a^2A_2^2\sin^2\theta\right)}{B^2}dt^2 
-\frac{2a\left[\left(r^2+a^2\right)A_1A_2-\Delta\right]\tilde{\Sigma}\sin^2\theta}{B^2}dtd\phi \nonumber\\
&&
+\frac{\tilde{\Sigma}}{\Delta A_5}dr^2+\tilde{\Sigma} d\theta^2   
+\frac{\left[\left(r^2+a^2\right)^2A_1^2-a^2\Delta\sin^2\theta\right]\tilde{\Sigma}\sin^2\theta}{B^2}d\phi^2 \, ,
\end{eqnarray}
where $M$ is the black hole mass, $a = J/M$, $J$ is the black hole spin angular momentum, $\tilde{\Sigma} = \Sigma + f$, and
\begin{eqnarray}
\Sigma = r^2 + a^2 \cos^2\theta \, , \quad
\Delta = r^2 - 2 M r + a^2 \, , \quad
B = \left(r^2+a^2\right)A_1-a^2A_2\sin^2\theta \, .
\end{eqnarray}
The functions $f$, $A_1$, $A_2$, and $A_5$ are defined as
\begin{eqnarray}\label{eq-fa1a2a5}
&& f = \sum^\infty_{n=3} \epsilon_n \frac{M^n}{r^{n-2}} \, , \quad
A_1 = 1 + \sum^\infty_{n=3} \alpha_{1n} \left(\frac{M}{r}\right)^n \, , \nonumber\\
&& A_2 = 1 + \sum^\infty_{n=2} \alpha_{2n}\left(\frac{M}{r}\right)^n \, , \quad
A_5 = 1 + \sum^\infty_{n=2} \alpha_{5n}\left(\frac{M}{r}\right)^n \, ,
\end{eqnarray}
where $\{ \epsilon_n \}$, $\{ \alpha_{1n} \}$, $\{ \alpha_{2n} \}$, and $\{ \alpha_{5n} \}$ are four infinite sets of deformation parameters without constraints from the Newtonian limit and Solar System experiments.

Up to now, we have mainly used this metric in our work, and in particular its simplest version in which all the deformation parameters vanish with the exception of $\alpha_{13}$. This is because, for the moment, we have mainly focused our efforts to improve the astrophysical model of {\tt relxill\_nk} and {\tt nkbb}, paying less attention to the possibility of constraining different deformation parameters or specific theories of gravity. However, the technique is very general and we can potentially test any stationary and axisymmetric black hole solution with a metric in analytic form.


\section{Results and conclusions}

We have used {\tt relxill\_nk} and {\tt nkbb} to analyze reflection features and thermal spectra of accreting black holes with X-ray observations of \textsl{NuSTAR}, \textsl{Suzaku}, \textsl{XMM-Newton}, and \textsl{RXTE}. We have studied either stellar-mass and supermassive black holes. All our results are consistent with the hypothesis that the spacetime metric around astrophysical black holes is described by the Kerr solution of General Relativity.

\begin{figure}[t]
\centerline{\includegraphics[width=0.6\linewidth]{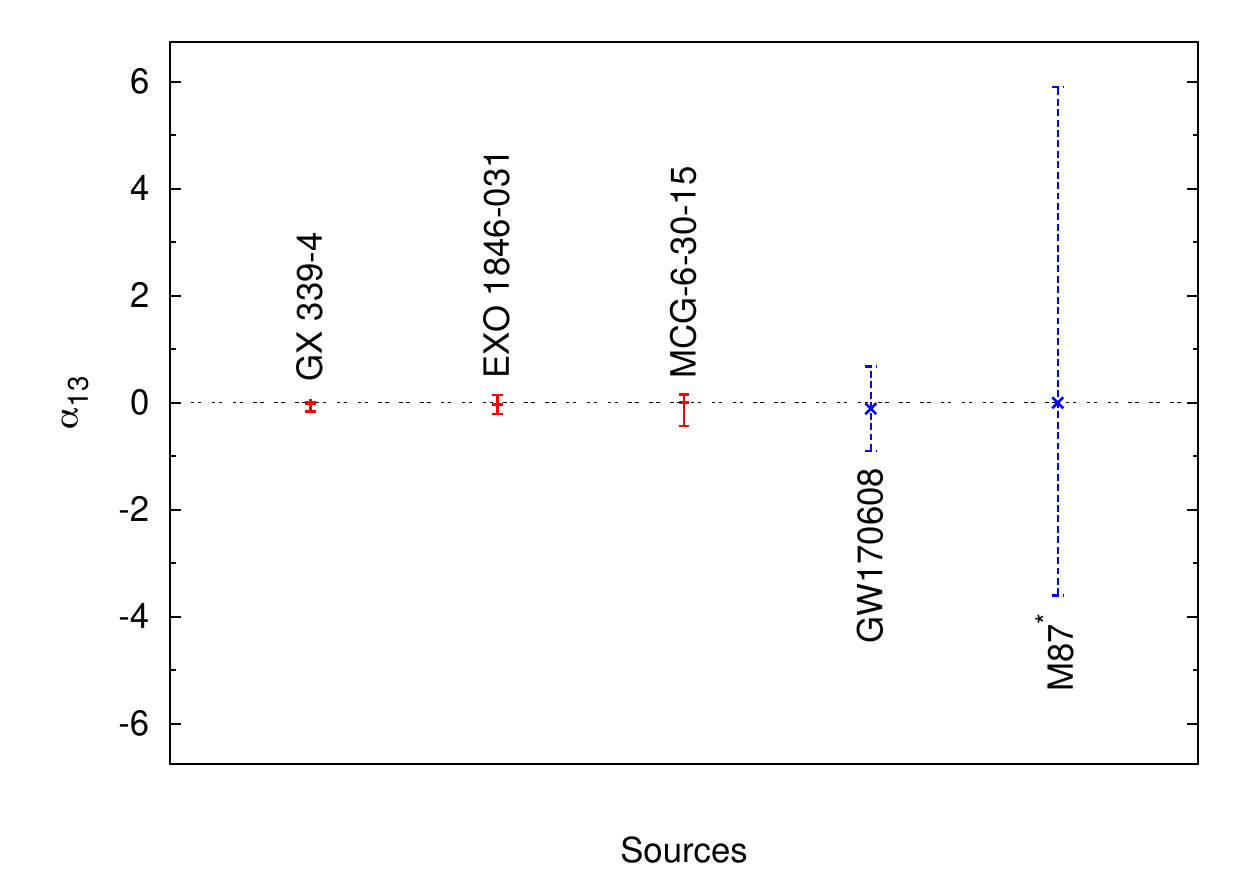}}
\caption[]{3-$\sigma$ measurements of the Johannsen deformation parameter $\alpha_{13}$ from the analysis of X-ray data of the stellar-mass black holes in GX~339--4~\cite{Tripathi:2020dni} and EXO~1846--031~\cite{Tripathi:2020yts} and of the supermassive black hole in MCG--06--30--15~\cite{Tripathi:2018lhx}. For EXO~1846--031 and MCG--06--30--15, the constraint on $\alpha_{13}$ is obtained with {\tt relxill\_nk}, while for GX~339--4 the constraint is inferred by the simultaneous use of {\tt relxill\_nk} and {\tt nkbb}. The figure also shows the 3-$\sigma$ measurement of $\alpha_{13}$ from GW170608~\cite{Cardenas-Avendano:2019zxd} and the 1-$\sigma$ measurement from the image of M87*~\cite{Psaltis:2020lvx}.}
\label{f-summary}
\end{figure}

The state-of-the-art in the field is our simultaneous analysis of reflection features and thermal spectrum of \textsl{NuSTAR} and \textsl{Swift} data of GX~339--4~\cite{Tripathi:2020dni}, which provides the most stringent and robust constraint on the Johannsen deformation parameter $\alpha_{13}$. Constraints on $\alpha_{13}$ with \textsl{NuSTAR} data of Galactic black holes are reported in Ref.~\cite{Tripathi:2020yts}. In the case of supermassive black holes, the most stringent and robust constraint on $\alpha_{13}$ was obtained from the analysis with {\tt relxill\_nk} of simultaneous observations \textsl{NuSTAR} and \textsl{XMM-Newton} of the galaxy MCG--06--30--15~\cite{Tripathi:2018lhx}. Fig.~\ref{f-summary} shows our best measurements of $\alpha_{13}$ with {\tt relxill\_nk} or with the simultaneous use of {\tt relxill\_nk} and {\tt nkbb} (red measurements). Our results are compared with the most stringent constraints with gravitational waves using LIGO/Virgo data (GW170608, blue measurement)~\cite{Cardenas-Avendano:2019zxd} and with black hole imaging from EHT data (M87*, blue measurement)~\cite{Psaltis:2020lvx}.

{\tt relxill\_nk} has been employed even to constrain specific gravity models, and we have tested conformal gravity~\cite{Zhou:2018bxk}, Kaluza-Klein gravity~\cite{Zhu:2020cfn}, asymptotically safe gravity~\cite{Zhou:2020eth}, and Einstein-Maxwell dilaton-axion gravity~\cite{Tripathi:2021rwb}. Our measurements of the spacetime around black holes are all consistent with the Kerr metric and we do not see any indication of possible new physics.

For the future, we want to further improve the theoretical models behind {\tt relxill\_nk} and {\tt nkbb} to reduce the systematic uncertainties and have more accurate measurements. In particular, we think that an important development in {\tt relxill\_nk} is the inclusion of the returning radiation, namely of the radiation emitted by the disk and returning to the disk because of the strong light bending near the black hole. Such an effect is currently ignored. More precise measurements of the spacetime metric will be possible with the next generation of X-ray missions. \textsl{XRISM} and \textsl{Athena} promise to have exceptional energy resolution in the iron line region. With \textsl{eXTP}, we can potentially have simultaneous spectral, timing, and polarimetric analyses.


\section*{Acknowledgments}

This work was supported by the Innovation Program of the Shanghai Municipal Education Commission, Grant No.~2019-01-07-00-07-E00035, the National Natural Science Foundation of China (NSFC), Grant No.~11973019, and Fudan University, Grant No.~JIH1512604.


\end{document}